\documentstyle[11pt,epsf]{article}
\def\Ex#1{\langle#1\rangle}
\def\qed{\nobreak\kern 1em \vrule height .5em width .5em depth 0em}
\def\vbar{\mathchoice{\vrule height6.3ptdepth-.5ptwidth.8pt\kern-.8pt}
   {\vrule height6.3ptdepth-.5ptwidth.8pt\kern-.8pt}
   {\vrule height4.1ptdepth-.35ptwidth.6pt\kern-.6pt}
   {\vrule height3.1ptdepth-.25ptwidth.5pt\kern-.5pt}}

\let\wt\widetilde
\def\date
   {\noindent Date: \today \par
    \medskip}

\def\ket#1{| #1 \rangle}
\def\b{\begin{eqnarray}}
\def\e{\end{eqnarray}}
\def\S{{\cal S}}
\def\p{\phi}
\def\pd{\phi^\dagger}
\def\pb{\overline{\phi}}
\def\t{\int_0^t dt}
\def\x{\int d^dx}
\def\wt#1{\widetilde{#1}}
\newcount\subno      
\newcount\subsubno   
\newcount\appno      
\newcount\appeqno    
\newcount\tableno    
\newcount\figureno   
\def\nsection#1
   {\vskip0pt plus.1\vsize\penalty-250
    \vskip0pt plus-.1\vsize\vskip24pt plus12pt minus6pt
    \subno=0 \subsubno=0
    \addtocounter{section}{1}
\renewcommand{\thesection}{\Roman{section}}
  {\small  \noindent {\bf \thesection. #1\par}}
    \noindent}
\def\nsubsecnn#1
   {\vskip-\lastskip
    \vskip12pt plus12pt minus6pt
    {\noindent {\bf #1\par}}
    \noindent}
\begin{document}
\setcounter{page}{1}
\pagestyle{plain}
\setcounter{equation}{0}
%
%
\ \\[12mm]
\begin{center}
    {\bf THE REACTION PROCESS $A+A\rightarrow O$ IN SINAI DISORDER\\[10mm]}
\end{center}
\begin{center}
\normalsize
M. J. E. Richardson$^{(1)}$ and John Cardy$^{(2)}$\\[5mm]
(1) Physics of Complex Systems, Weizmann Institute of Science,\\ Rehovot, 76200, Israel\\
(2) Department of Theoretical Physics, University of Oxford, \\ 1 Keble Road, Oxford, OX1 3NP, UK.\\
and All Souls College, Oxford.
\\[3mm] 
\end{center}
{\bf Abstract:} 
The single-species reaction-diffusion process $A+A\rightarrow O$ is examined in the presence of an uncorrelated, quenched random velocity field. Utilising a field-theoretic approach, we find that in two dimensions and below the density decay is altered from the case of purely diffusing reactants. In two-dimensions the density amplitude is reduced in the presence of weak disorder, yielding the interesting result that Sinai disorder can cause reactions to occur at an {\it increased} rate. This is in contrast to the case of long-range correlated disorder, where it was shown that the reaction becomes sub-diffusion limited.  However, when written in terms of the microscopic diffusion constant it is seen that increasing the disorder has the effect of reducing the rate of the reaction. Below two dimensions, the effect of Sinai disorder is much more severe and the reaction is shown to become sub-diffusion limited. Although there is no universal amplitude for the time-dependence of
the density, it is universal when expressed in terms of the
disorder-averaged diffusion length. The appropriate amplitude is
calculated to one-loop order.

\date 
\rule{7cm}{0.2mm}
\begin{flushleft}
\parbox[t]{3.5cm}{\bf Key words:} reaction system, disorder, non-equilibrium physics, statistical mechanics
\parbox[t]{12.5cm}{ }
\\[2mm]
\parbox[t]{3.5cm}{\bf PACS numbers:}  82.30.-b  05.70.Ln  02.50.Ey  64.60.Ht  
\\[2mm]
\parbox[t]{3.5cm}{\bf Short title:} Reactions in Sinai disorder
\end{flushleft}
\normalsize
\thispagestyle{empty}
\mbox{}
\pagestyle{plain}
%
%
%
\newpage
\setcounter{page}{1}
\setcounter{equation}{0}
\section{Introduction}
The dynamical, many-body problem of diffusing and reacting chemicals provides an ideal testing ground for the many methods of non-equilibrium physics. Such reaction systems occur in nature in a wide variety of guises, from conventional chemical reactions to more exotic processes such as domain coarsening in magnetic systems \cite{Bray},\cite{Derr} and exciton annihilation in crystals \cite{TMMC}. When an equilibrium has been reached in a reversible reaction, the methods of statistical mechanics can be used to give the appropriate reactant-density ratios. However, in the decay to equilibrium \cite{CARDY} and in irreversible processes no such unified approach yet exists. Nevertheless, over the last few decades a great many studies have been made utilising numerical, exact and renormalisation-group techniques, and the time-dependence of reaction systems has been found to be remarkably rich. It has been shown that the collective behaviour of the reaction system is rather sensitive to the statistical properties of a single reactant's motion, with a basic example given by the case of the single-species diffusive reaction $A+A\rightarrow O$. This is because the re-entrancy of random walks in two dimensions and below alters the density decay from the simple mean-field prediction. In these lower dimensions, the density loses all dependence on the microscopic reaction rate and becomes a universal function of only the diffusion length: the reaction is {\it diffusion limited}. In this paper we examine the behaviour of the $A+A\rightarrow O$ reaction process in the presence of an uncorrelated, quenched random velocity field. Before going on to describe our results and method, we review two cases, first the pure reaction-diffusion process, and second some studies of the reaction process in {\it correlated} long-range potential disorder.

The mutually annihilating random walk (MARW) $A+A\rightarrow O$ is a fundamental theoretical model in the study of reaction systems. It describes the process whereby diffusing $A$ particles may react pairwise, at a rate $\lambda$ on contact. The MARW and coalescing random walk $A+A\rightarrow A$ are members of the same universality class \cite{OLD}, and therefore the results below are also valid for the coalescence process (albeit with trivial changes in the prefactors). The basic starting point in the analysis of such systems is the mean-field or rate equation. This corresponds to writing a self-consistent equation for the average reactant density $\overline{n}$ as a function of $t$, that ignores all spatial correlations (and more seriously {\it anti-correlations}),
\b
\frac{\partial \overline{n}}{\partial t}=D\nabla^2\overline{n}-2\lambda\overline{n}^2,&\mbox{ with the late time result }&\overline{n}\simeq\frac{1}{2\lambda t} \label{mf}
\e
where we have introduced the diffusion constant $D$ and the reaction rate is $\lambda$. Neglecting the effect of correlations is equivalent to assuming that the reactants remain well mixed throughout the reaction process. However, due to the statistics of random walks in two dimensions and below, simple diffusion of particles itself is not sufficiently fast to maintain a well-mixed state. For this reason, the mean-field result (\ref{mf}) loses its validity in two dimensions and below. In these lower spatial dimensions the re-entrancy of random walks means that reactants come into contact many times, each time providing an opportunity for a reaction to take place. This implies that even a small reaction rate $\lambda$ does not limit the global rate of reaction. Therefore, in two dimensions and below the reaction becomes diffusion limited with the density decays
\b
\hspace{0.5cm}n=\frac{\log(Dt)}{8\pi Dt}\hspace{0.5cm}\mbox{ for $d=2$,}\hspace{1cm}n\simeq\frac{{\cal A}_d}{(Dt)^{d/2}}\hspace{0.5cm}\mbox{ for $d<2$} \label{PDC}
\e
where ${\cal A}_d$ is a universal amplitude, a function only of the dimension $d$. This amplitude has been calculated via an $\epsilon$ expansion for $d=2-\epsilon$ dimensions \cite{BPL} (also given in equation (\ref{Gloop})) and the exact result ${\cal A}_1=(8\pi)^{-1/2}$ for one dimension can be found in reference \cite{LUSH}.

Over the last few years there has been an increased interest in the behaviour of reaction systems with reactants that perform motion different from pure diffusion. As well as ballistic gas-phase reactions \cite{ELSKINS}--\cite{RICHARDSON}, studies have appeared with reactants that perform diffusion in the presence of turbulence \cite{PDTURB1} \cite{PDTURB2} and also in quenched random velocity fields. Some basic categories for the statistical properties of these random velocity fields have been identified, see for example \cite{KLY} \cite{F2}. In particular, a distinction can be made between uncorrelated {\it Sinai} disorder and the long-range correlated potential disorder, with the momentum-space correlator $\gamma/k^2$. Studies have been made of the behaviour of the reaction front in the segregated two-species $A+B\rightarrow O$ reaction with various forms of Sinai disorder \cite{KO}--\cite{KT} and correlated potential disorder \cite{PDAB1} \cite{PDAB2}. A  comprehensive study has been made of the $A+A\rightarrow O$ scheme with correlated potential disorder \cite{AAPD} \cite{CD} in two spatial dimensions and solutions also exist for this single-species reaction with random barriers and random traps \cite{SCHUTZ}. Furthermore, the specific case of single-species reactions in Sinai disorder was recently examined in one dimension in the context of aging phenomena, with the persistence exponent derived \cite{FDM}. In the two-dimensional case of $A+A\rightarrow O$ in potential disorder, it was found that the reaction process becomes {\it sub-diffusion} limited: the reaction rate is lower than the MARW. The form for the density with this potential disorder, which should be compared with the two-dimensional MARW in equation (\ref{PDC}), was found to be
\b
n\simeq\frac{1}{\lambda^* t^{1-\delta}}&\mbox{ with }&\delta=\left[1+\frac{8\pi}{\beta^2\gamma}\right]^{-1}>0,\;\;\;\lambda^*=3D\beta^2\gamma \label{LRD}
\e
where $\beta^2\gamma$ measures the disorder strength, $\delta$ is a non-universal exponent, and $\lambda^*$ is an effective reaction constant. The interpretation was that the long-range disorder produces potential traps on all length scales which, after some time, will contain at most one reactant. Reactions can then only occur when the trapped reactants move between traps. However, at the same time, as the reactants explore the landscape, they get caught by increasingly deep traps, leading to sub-diffusive motion.

In this paper, we present results from a study of the single-species $A+A\rightarrow O$ reaction process in the presence of an uncorrelated, quenched random velocity field in dimension two and below. Despite the lack of long-range correlations the kinetic behaviour of diffusing particles is changed, as can be seen in the behaviour of the diffusion-length squared $\Ex{r^2}$. In two dimensions, it is known from a renormalisation-group (RG) treatment that this length is altered by the presence of a logarithm \cite{F1}. Below two dimensions, $d=2-\epsilon$, the RG gives the two-loop result for the dynamic exponent of $z=2+2\epsilon^2+O(\epsilon^3)$ and hence the motion is sub-diffusive. The time-dependence in $d=1$ is also known \cite{SINAI}, giving the following behaviour for $\Ex{r^2}$ as a function of dimension
\b
\Ex{r^2}&\propto&[\log(Dt)]^4  \hspace{0.5cm} \mbox{ for $d=1$}\nonumber \\
\Ex{r^2}&\propto& (Dt)^{2/z}  \hspace{0.9cm} \mbox{ for $d<2$} \nonumber \\
\Ex{r^2}&\simeq&4D_Rt\left[1+\frac{4}{\log(t)}+O\left(\frac{1}{\log^2(t)}\right)\right] \hspace{0.5cm} \mbox{ for $d=2$ in weak disorder} \label{DL}
\e
where, for comparison, the result for pure diffusion is $\Ex{r^2}=2dDt$, and $D_R$ in equation (\ref{DL}) is the effective, measured diffusion constant in the late-time limit. 

We will show that below two dimensions, a universal form similar to the pure-diffusion $d<2$ result in (\ref{PDC}) with $n$ a function of the time $t$ cannot be found. This is due to the changed dynamic exponent $z$ which requires a dimensionful amplitude that must be a function of the disorder strength or the reaction rate. Though, it is reasonable to study the density as a function of time, or more specifically as a function of the length $Dt$, it is not the appropriate length scale for the reaction-diffusion problem in the presence of disorder. The natural length to use is the disorder-averaged diffusion length. By rewriting the density decay as a function of the scale $\Ex{r^2}$ a fully universal relation, similar to (\ref{PDC}) can again be found
\b
n\simeq\frac{{\cal B}_d}{\Ex{r^2}^{d/2}}&\mbox{ with }& {\cal B}_d=\left[\frac{1}{3\pi\epsilon}+\frac{2\log(128\pi)-11}{12\pi}+O(\epsilon)  \right]\;\mbox{ for $d=2-\epsilon$}. \nonumber
\e

The effect of Sinai disorder in two dimensions is not strong, and the alteration to the diffusion length (\ref{DL}) is not leading order. Nevertheless, we find the interesting result that a reaction process occurring in this disorder has a decay rate with a different {\it leading-order} amplitude from the MARW,
\b
n&=&\frac{\log(t)}{24\pi D_Rt}+O\left(t^{-1}\right). \nonumber
\e
In fact for weak disorder the reactions occur {\it faster} than in the MARW, contrary to the effects seen in the case of long-range potential disorder. Nevertheless, when written in terms of the diffusion constant of the underlying lattice model (to be described below) it will be shown that the disorder strength increases the density's amplitude. These effects come from two competing  terms: the disorder-renormalisation of the reaction term that increases the rate of reaction and the disorder-renormalisation of the propagator that decreases the rate of reaction. 

In the rest of this paper we describe how these results were derived in more detail. In section (2) we introduce the model and describe some of the steps taken in the field-theoretic analysis of the model. In particular, the relation to existing theories, of diffusion in Sinai disorder \cite{F1} and reaction with pure diffusion \cite{OLD} \cite{BPL} are discussed. The fixed point structure of the renormalised parameters is found and a perturbation expansion for the density, valid at early times, is obtained for $d\leq2$. In section (3) we obtain a Callan-Symazik (CS) equation for the density as a function of time and show that no universal functional form can be found for $d<2$. However, by re-expressing the density in terms of the disorder-averaged diffusion length a universal form is obtained and the amplitude calculated to one-loop order. The behaviour at the upper-critical dimension $d_c=2$ is then examined and the density as function of the disorder strength analysed. Finally, we close in section (4) with a discussion of the results obtained.

\section{The model and method}
In this section, we introduce the model to be studied and also describe some of the steps taken to achieve its representation in field-theoretic form. The method used is standard and we only dwell on details that are different from systems previously studied. After the model has been defined it is written in the language of second quantisation, which in turn allows a mapping to a path-integral formulation. An average over all possible realisations of the random velocity field can be taken at this point, to produce a weighting function (an action) that gives disorder-averaged correlation functions. This bare action is then regularised and finally used to calculate a perturbation expansion for the early-time, disorder-averaged reactant density.

The model is defined on an infinite $d$-dimensional hypercubic lattice with a lattice spacing of unity. Each site $i$ of this lattice contains $n_i$ particles where $n_i$ can take the values $0,1,2\cdots$. The quenched disorder in the diffusion rates is modeled by particles hopping independently from a lattice site $i$ to a neighbouring site $e$ at a fixed rate $p_{i\rightarrow e}$. The rates $\{p\}$ are random and contain no long-range correlations. Reactions can occur if there are two or more particles, $n_i\geq2$, on a lattice site. This happens at a rate $\lambda n_i(n_i-1)$, where $\lambda$ is the on-site reaction rate, reducing the number of particles on that site by 2.

The field-theoretic description is obtained by  writing a {\it master equation} that describes the time-dependent flow of probability between microstates. It is convenient to write this equation in the language of bosonic operators. Given that the set of occupation numbers $\{n_i\}$ defines a microstate of the system, the probability that the system is in such a microstate will be written $P(\{n_i\})$. The master equation is $\partial_t\ket{\psi(t)}=-{\cal H}\ket{\psi(t)}$, where the probability state vector $\ket{\psi(t)}$ and evolution operator ${\cal H}$ are
\b
\ket{\psi(t)}&=&\sum_{\{n_i\}}P(\{n_i\})\prod_j(a_j^\dagger)^{n_j}\ket{0} \nonumber \\
{\cal H}&=& \sum_i\left[D\sum_e\left(p_{i\rightarrow e}a^\dagger_ia_i-p_{e\rightarrow i}a^\dagger_ia_e\right)-\lambda\left(1-(a^\dagger_i)^2\right)a_i^2\right]. \label{H}
\e	
This algebraic description can now be converted to a field theory by using the coherent-state formalism. Observables, like the expected density $\Ex{n_j(t,\{p\})}$ at site $j$ at time $t$ for a given realisation of the disorder $\{p\}$ can be written as a path integration with respect to an action $\S_p$
\b
\Ex{n_j(t,\{p\})}&=&\int\prod_i\left[{\cal D}\p_i{\cal D}\pd_i\right]\;\p_j\;\exp{\left(-\S_p\right)}. \nonumber
\e
The integration is over the complex fields $\p,\pd$ and the action $\S_p$ derived from equation (\ref{H}) is
\b
\S_p\!\!&=&\!\!\sum_i\left[-\p_i(t)+\t\left(\pd_i\partial_t\p_i+\pd_i\sum_e\left(p_{i\rightarrow e}\p_i-p_{e\rightarrow i}\p_e\right)-\lambda(1-(\pd_i)^2)\p^2_i\right)-n_0\p_i(0) \right]. \nonumber
\e
It is convenient to shift the field $\pd$ by its classical value $\pd=\pb+1$ and take the continuum limit in space. The action $\S(\vec{V})$ thus obtained is naturally split into four parts $\S_D+\S_{\vec{V}}+\S_R+\S_{n_0}$: the diffusive, disorder, reaction and initial conditions. For the moment let us examine the diffusive and disorder parts
\b
\S_D&=&\t\x\left(\pb\partial_t\p-\pb\nabla^2\left(D(x)\p\right)\right) \nonumber \\
\S_{\vec{V}}&=&-\t\x \pb\nabla\cdot\left(\vec{V}(x)\p\right). \nonumber
\e
The disorder appears in both the diffusion constant and in a random velocity vector-field $\vec{V}(x)$. However, as can be checked under the RG, the disordered, spatially varying component of $D(x)$ is irrelevant in the technical sense, and therefore we consider just the case of a uniform diffusion field $D(x)=D$. Furthermore, though we have used a lattice model in the derivation of the continuum field theory, it is not necessary that some of the restrictions of the lattice formulation are passed to the continuum theory. In particular, for a walker on a lattice with a bias $V$ there is a minimum dispersion $D=V/2$. In the continuum, there is no reason to impose such a restriction, and therefore we treat the magnitude of the diffusion constant and $V(x)$ as fully independent quantities.

The velocity vector-field $\vec{V}(x)$ is taken to be a Gaussian random variable with the correlator $\Ex{V^{\alpha}(x)V^\beta(y)}=\Delta\delta_{\alpha,\beta}\delta(x-y)$, i.e. there are no long-range correlations. An average can be performed on the component $\S_{\vec{V}}$ with respect to this field, to produce an action that gives disorder-averaged correlation functions,
\b
\exp\left(-\S_\Delta\right)=\int{\cal D}\vec{V}\left[\;\exp\left(-\frac{1}{2\Delta}\x(\vec{V}(x))^2\right)\exp\left(-\S_{\vec{V}}\right)\right]. \nonumber
\e
This integral can be performed to give the disorder-averaged bare action $\S_0=\S_D+\S_\Delta+\S_R+\S_{n_0}$, which when suitably regularised, can be used for calculations. The various components of this action are
\b
\S_D&=&\t\x\pb\left(\partial_t-D_0\nabla^2\right)\p \label{action} \\
\S_\Delta&=&\frac{\Delta_0}{2}\x\left(\t\p\nabla\pb \right)\cdot\left(\t\p\nabla\pb \right) \nonumber \\
\S_R&=& 2\lambda_0\t\x\pb\p^2+\lambda_0\t\x\pb^2\p^2 \nonumber \\
\S_{n_0}&=&-n_0\x\t\pb\delta(t).\nonumber
\e
Where we have written $D_0$, $\Delta_0$ and $\lambda_0$ with subscripts to stress that they are bare quantities, but $n_0$ represents the initial density at $t=0$. It is seen that the action is a combination of that found in \cite{F1} for diffusion in Sinai disorder (though with a trivial change of field variables) and that of the purely diffusive reaction process \cite{BPL}. Following the notation in previous works, the diagrams for the vertices are given in figure (1). The two reaction vertices figure (1a), renormalise identically and hence we do not introduce a separate reaction parameter for each vertex. It should also be noted that the disorder vertex figure (1b), when considered in momentum space, is proportional to the scalar product of the momentum flowing through the two outgoing $\pb$ fields. Finally, as with both the pure reaction and pure disordered-diffusion theories individually, the upper-critical dimension for the hybridised theory is $d_c=2$.

Because taking the continuum limit in space introduces unphysical divergences, the theory must be rendered finite before calculations can proceed. This was achieved by dimensional regularisation in $d=2-\epsilon$ dimensions in the absence of the $\S_{n_0}$ term. In this theory there is no field renormalisation, in fact only the diffusion constant, disorder strength and reaction rate are renormalised. The following renormalised diffusion constant $D$ and dimensionless interaction parameters $g$ and $h$ are now introduced
$$
Z_DD=D_0 \hspace{1cm} Z_ggD\mu^\epsilon=Z_g\lambda=\lambda_0 \hspace{1cm} Z_hhD^2\mu^\epsilon=Z_h\Delta=\Delta_0.
$$
It can be shown that the disorder vertex and propagator are not renormalised by the reaction vertices, and therefore we can use the results found previously \cite{F1} for $Z_h$ at the one-loop level and $Z_D$ at the two-loop level. In fact the only new diagrams that need be considered are the dressings of the the reaction strength by the disorder. As can be checked, only one such diagram, see figure (1), diverges in $d=2$. This contribution is combined with the previously known result for the reaction-reaction renormalisation to give the following set of $Z$ factors
$$
Z_D=1+\frac{h^2}{(4\pi)^2\epsilon}+\ldots\hspace{1cm}Z_g=1+\frac{(g-h)}{2\pi\epsilon}+\ldots\hspace{1cm}Z_h=1+\frac{h}{4\pi\epsilon}+\ldots. 
$$
The flow function for the diffusion constant and the beta functions can now be evaluated
\b
\varrho=\frac{\partial \log(D)}{\partial \log(\mu)}&=&\frac{2h^2}{(4\pi)^2} +O(h^3)\nonumber \\
\beta_g=\frac{\partial g}{\partial \log(\mu)}&=&\frac{g}{2\pi}\left(g-h-2\pi\epsilon\right)+O(g^3,g^2h,gh^2) \nonumber \\
\beta_h=\frac{\partial h}{\partial \log(\mu)}&=&\frac{h}{4\pi}\left(h-4\pi\epsilon\right)+O(h^3). \nonumber 
\e
The fixed point structure for $h$ is of course unchanged by the presence of the reaction vertex, with $h^*=4\pi\epsilon+O(\epsilon^2)$ and the dynamic exponent remains $z=2+\varrho^*=2+2\epsilon^2+O(\epsilon^3)$. However, the fixed point of the renormalised reaction strength is now shifted to the larger value of $g^*=6\pi\epsilon+O(\epsilon^2)$ in the presence of Sinai disorder (the value with no disorder is $g^*=2\pi\epsilon+O(\epsilon^2)$). 

\begin{figure}
\begin{minipage}[b]{8cm}
\epsfxsize 8 cm
\epsfysize 8 cm
\epsfbox{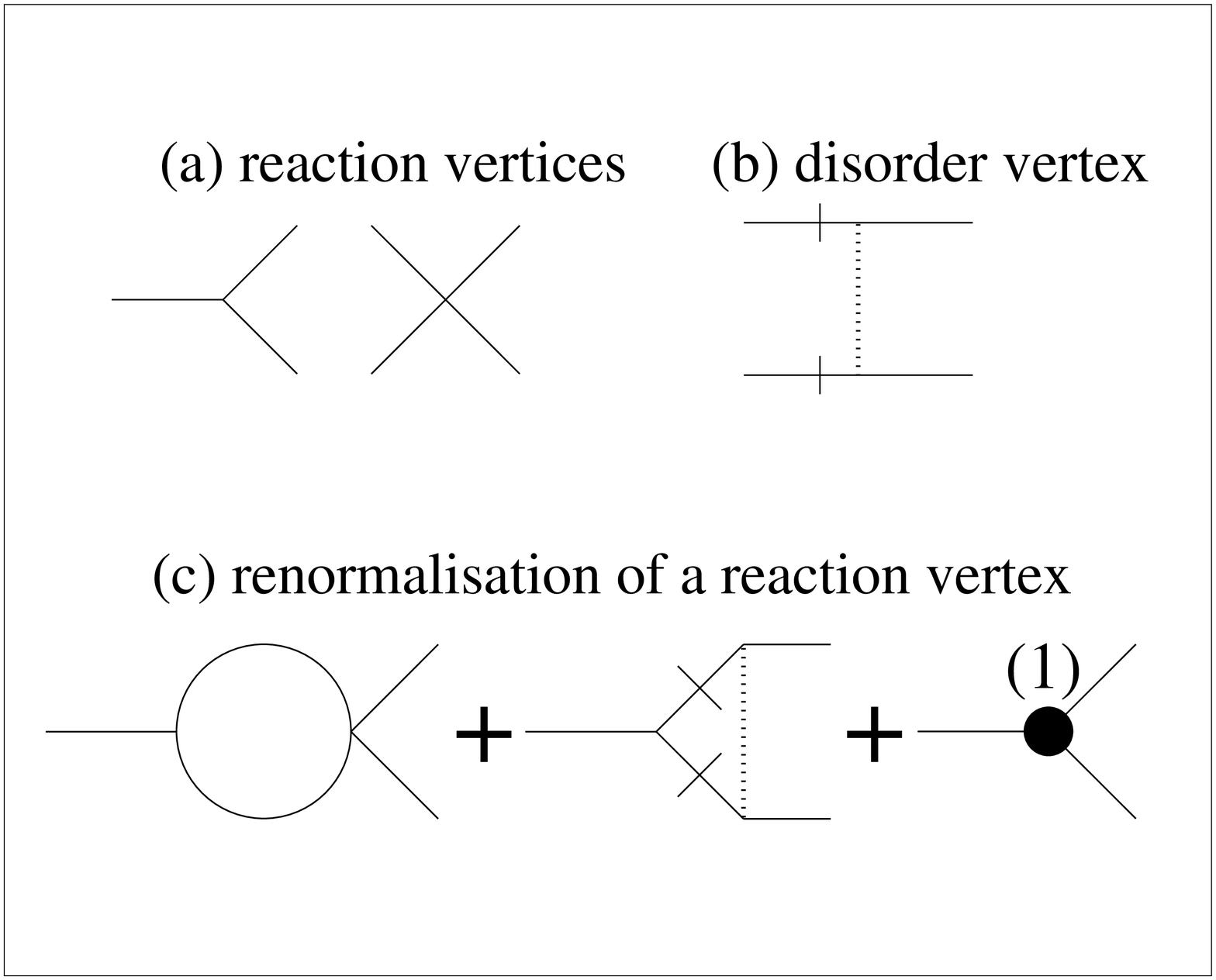}
\end{minipage}
\begin{minipage}[b]{8cm}
\epsfxsize 8 cm
\epsfysize 8 cm
\epsfbox{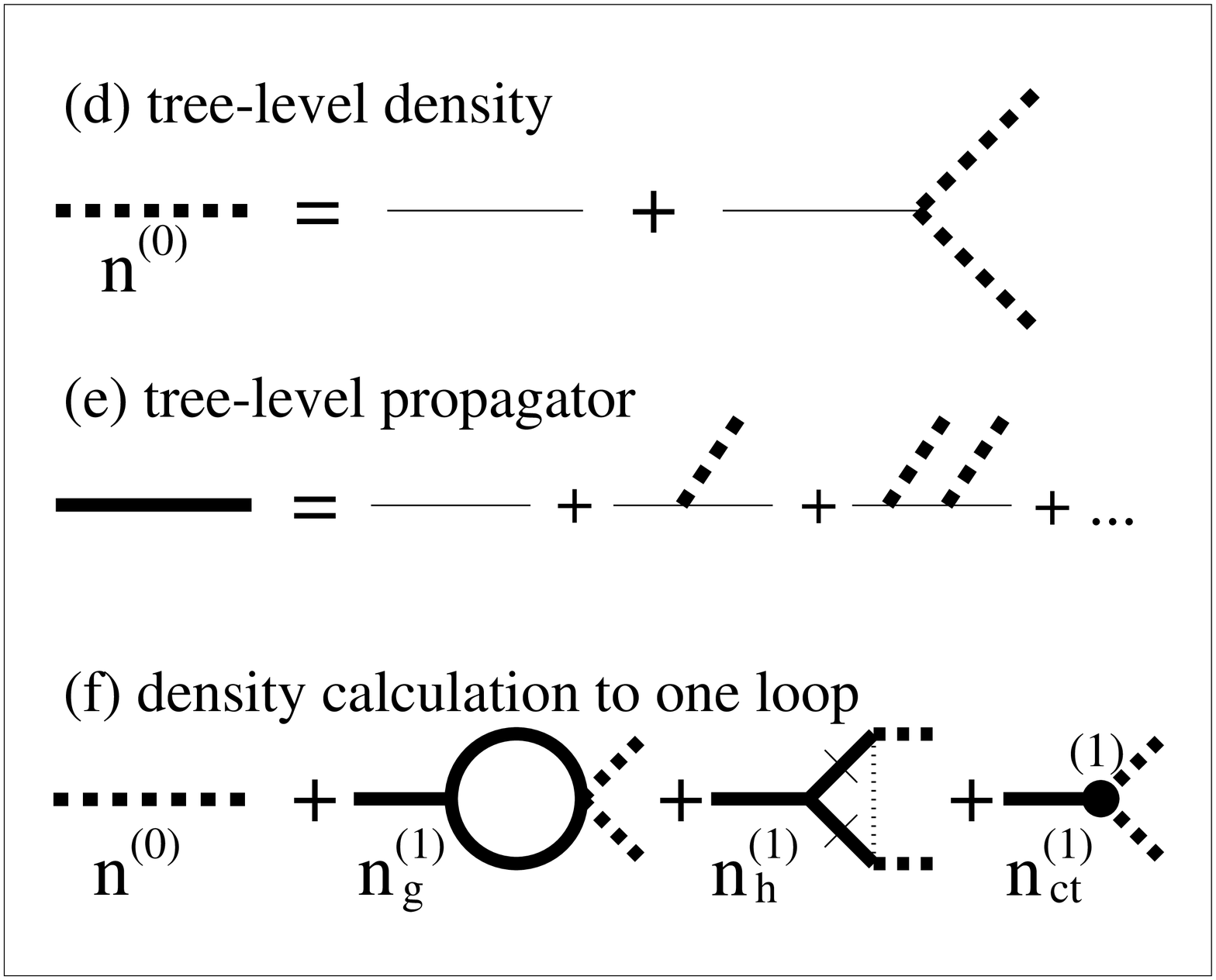}
\end{minipage}
\caption{The various quantities expressed as combinations of the vertices (1a) and (1b). The requirement that the sum (1c) is finite gives the one-loop contribution to the renormalisation constant $Z_g$. Diagram (1d) is the mean-field density given in equation (7). Diagram (1e) is the propagator dressed by the tree-level reaction interaction and is used in the one-loop contribution to the density calculation shown in diagram (1f).}
\end{figure}

\nsubsecnn{The perturbative density}
With the theory now regularised, observables such as the reactant density $n(t)$ at an early time $t$ can be calculated perturbatively
\b
n(t)&=&\int\left[{\cal D}\p{\cal D}\pb\right]\;\p\;\exp{\left(-\S-\S_{ct}\right)} \nonumber
\e
where the bare action has been rewritten in terms of the renormalised action and the appropriate counter term action $\S_0=\S+\S_{ct}$. The initial conditions are included and the renormalised action is equivalent in form to the bare action written above, but with the replacements $D_0\rightarrow D$, $\Delta_0\rightarrow hD^2\mu^\epsilon$ and $\lambda_0\rightarrow gD\mu^\epsilon$. The early-time density calculation is performed as a loop expansion of equation (2) taken to the appropriate order of the small parameter $\epsilon=2-d$. An expansion is also made in inverse powers of the initial density $n_0$ with only the leading-order, $n_0$ independent term retained (the reason for this limit of large initial density will be given in the next section). 

The tree-level or mean-field result shown in figure (1d) is is unchanged from the case of the MARW, equation (\ref{mf})
\b
n^{(0)}&=&\frac{1}{2g\mu^\epsilon Dt}. \label{tree}
\e
However, at the one-loop level a new diagram appears, representing two particles interacting with the same disordered region at different times, and eventually annihilating. This addition, the third shown in figure (1f), is equal to
\b
n_h^{(1)}&=& -\frac{1}{(Dt)^{d/2}}\frac{h}{g}\left[\frac{1}{4\pi\epsilon}+\frac{2\log(\pi)-3}{16\pi}\right]. \nonumber
\e
To produce the full, regularised one-loop contribution, this result is combined with the reaction-reaction contribution $n_g^{(1)}$, which is also common to \cite{BPL}, and the appropriate counter term $n_{ct}^{(1)}$
\b
n^{(1)}_g= \frac{1}{(Dt)^{d/2}}\left[\frac{1}{4\pi\epsilon}+\frac{2\log(8\pi)-5}{16\pi}\right]&\hspace{5mm}&n^{(1)}_{ct}=\frac{h-g}{4\pi\epsilon g\mu^\epsilon Dt}. \label{Gloop}
\e
The full result can be expanded to finite order in $\epsilon$ to produce the full one-loop perturbative result
\b
n^{(1)}&=&\frac{1}{16\pi(Dt)^{d/2}}\left[2\log(8\pi)-5-\frac{h}{g}\left(2\log(\pi)-3\right)\right]+O(\epsilon). \label{1loop}
\e

\section{The reactant density}
In this section, the late-time behaviour of the reactant density will be examined. The perturbative density, has already been derived and is given by the sum of equations (\ref{tree}) and (\ref{1loop}). Now a Callan-Symanzik (CS) equation will be used to relate this perturbative density to the required late-time, non-perturbative density. First, the differential CS equation will be obtained and solved as a function of time $t$ and the case of dimensions less than two will then be examined. It will be shown that no universal relation giving the density as a function of time can be obtained. However, by trading the time dependence in the CS equation for dependence on the disorder-averaged diffusion length, a fully universal relation between this length scale and the reaction decay rate will be obtained. The behaviour at the upper-critical dimension is then examined. It will be shown that the reactant density in two dimensions, written in terms of the measured, late-time diffusion constant, decays at a {\it faster} rate than the case of pure diffusion-reaction. However, written in terms of the diffusion constant of the underlying lattice model, it is shown that as the disorder strength increases the reaction rate decreases.

\nsubsecnn{A Callan-Symanzik equation for the density} 
All physical quantities must be independent from the arbitrary $\mu$ that was introduced in section 2 in the definition of the dimensionless interaction strengths $g$ and $h$. Using this fact, and also dimensional analysis, the following differential equation can be written for the density $n$
\b
\left[(2+\varrho)\frac{\partial}{\partial\log(Dt)}+\beta_g\frac{\partial}{\partial g}+\beta_h\frac{\partial}{\partial h}-d\frac{\partial}{\partial \log(n_0)}+d\right]n&=&0 \label{CS}
\e
where the interaction dependent flow functions $\varrho$, $\beta_g$ and $\beta_h$ are given in the previous section. This can now be solved in the standard way by writing equation (\ref{CS}) as a complete differential with respect to a scaling variable $s$, thus
\b
\left[\frac{d}{d\log(s)}+d\right]\wt{n}&=&0. \nonumber 
\e
where $\wt{n}=n(s)$ is the value of the density at some scale. Following the notation used in \cite{BPL} we denote early-time quantities  $X(s)$ (at a scale $s$) as $\wt{X}$ and late-time quantities (at a scale $s=1$) simply as $X(1)=X$. The flow equations for the system parameters as a function of the scale $s$ are
\b
\frac{\partial \log(D\wt{t})}{\partial\log(s)}=2+\wt{\varrho}\;\;\;\;\frac{\partial \wt{g}}{\partial\log(s)}=\wt{\beta}_g,\;\;\;\;\frac{\partial \wt{h}}{\partial \log(s)}=\wt{\beta}_h,\!\!&&\;\;\frac{\partial \log(\wt{n_0})}{\partial \log(s)}=-d \label{flows}.
\e
and the equation relating the density at these two different scales $s=1$ and $s$ is
\b
n(Dt,g,h,n_0,\mu)&=&s^d \wt{n}(D\wt{t},\wt{g},\wt{h},\wt{n_0},\mu) \label{SOLUTION} \e
The procedure will now be to insert the perturbative results, (\ref{tree}) and (\ref{1loop}), suitably rewritten with $g\rightarrow\wt{g}$ etc, into the RHS of equation (\ref{SOLUTION}) and replace all arguments with the $s=1$ values using the solutions of equations (\ref{flows}). Before proceeding it should be noted that the solution of the flow equation for the initial density implies $\wt{n}_0=n_0/s^d$. As we will be interested in the late-time regime $s\rightarrow0$, this justifies the $1/\wt{n}_0$ expansion in the perturbative calculation in the previous section.

\nsubsecnn{Below two dimensions}
The behaviour of the reactant density for $d<2$ is now considered. From the flow equation (\ref{flows}) in $d=2-\epsilon$ it is seen that as $s\rightarrow0$ the quantities $\wt{g}$, $\wt{h}$ and $\wt{\varrho}$ approach their fixed-point values $6\pi\epsilon,4\pi\epsilon$ and $2\epsilon^2$ respectively. Thus, time varies with the scale $s$ as $s^z=\wt{t}/t$ where $z=2+2\epsilon^2$ at the two-loop level \cite{F1}. Using these results, the following form for the reactant density below two dimensions is found
\b
n(t)&\simeq&\frac{C}{(Dt)^{d/z}}. \label{NT}
\e
This form appears similar to the $d<2$ case in equation (\ref{PDC}) but actually lacks all the important universal features of that relation. By dimensional analysis, the prefactor $C$ must carry $(2-z)d/z$ units of length and must therefore be a function of $\lambda$, $\Delta$ or both. No such universal amplitude like ${\cal A}_d$ can be found that relates density to time for the case of reactions in a Sinai velocity field. Moreover, given that the dynamic exponent $z=2+2\epsilon^2+O(\epsilon^3)$ is itself an approximate quantity, there is little point attempting to derive the non-universal prefactor $C$. However, we will now show that, though there is no universality in a relation between the density and the length scale $Dt$ in the presence of disorder, a fully universal relation nevertheless exists in terms of the reaction density and a different length scale: the disorder-averaged diffusion length $\Ex{r^2}^{1/2}$ which is the typical distance a single reactant explores. This quantity can be shown to vary as $\Ex{r^2}\propto t^{2/z}$, also  with a non-universal amplitude. We now rewrite the CS equation (\ref{CS}) for the density, exchanging the dependence in time for a dependence in $\Ex{r^2}$
$$
\left[2\frac{\partial}{\partial\log\Ex{r^2}}+\beta_g\frac{\partial}{\partial g}+\beta_h\frac{\partial}{\partial h}-d\frac{\partial}{\partial \log(n_0)}+d\right]n=0.
$$
This is solved as above with the scaling $s^2=\wt{\Ex{r^2}}/\Ex{r^2}$. The perturbative density up to the one-loop level $\wt{n}^{(0)}+\wt{n}^{(1)}$ given in equations (\ref{tree}) and (\ref{1loop}) can also be rewritten with the substitution $2dDt=\wt{\Ex{r^2}}$ which is correct at this order. Combining these results gives the following late-time density expressed as a function of $\Ex{r^2}$
\b
n&=&\frac{1}{\Ex{r^2}^{d/2}}\left[\wt{\Ex{r^2}}^{d/2}n(\wt{\Ex{r^2}},g^*,h^*,\wt{n_0},\mu)\right] \label{RGr2} \nonumber \\
n&\simeq&\frac{1}{\Ex{r^2}^{d/2}}\left[\frac{1}{3\pi\epsilon}+\frac{2\log(128\pi)-11}{12\pi}+O(\epsilon)\right]. \label{ANSWER}
\e
Both $n$ and $\Ex{r^2}$ are non-universal functions of $t$, in as much as the disorder strength $\Delta$ enters explicitly. However, we have found a universal relation between them, independent of {\it all} system parameters except the dimension of space. Though the amplitude calculated at this order in an $\epsilon$ expansion is unlikely to give a good result for one dimension, the scaling relation $n\propto1/\Ex{r^2}^{d/2}$ is exact at all orders in perturbation theory. Therefore, in one dimension it is expected that the product $n(t)\Ex{r^2(t)}^{1/2}$ approaches a universal, constant value independent of the disorder strength, reaction rate, and initial density. This is in agreement with the result found in \cite{SCHUTZ} for the $A+A\rightarrow O$ with infinite reaction rate. Given the validity of a factorisation assumption made in \cite{SCHUTZ} for Sinai disorder, the expected exact result for one dimension would have given an amplitude $1/4\pi$ - equivalent to the MARW.

\nsubsecnn{Two dimensions}
We now consider the case of the reactant density in two dimensions, $d=2$. To obtain the asymptotic behaviour it will only be necessary to use the tree-level perturbative density $n^{(0)}$ rewritten in terms of the diffusion length $\wt{\Ex{r^2}}$. The relevant equations to be inserted into the scaling relation are
$$
\wt{n}^{(0)}= \frac{2}{\wt{g}\wt{\Ex{r^2}}}\hspace{1.5cm}\wt{g}\simeq-\frac{6\pi}{\log(s)}\hspace{1.5cm}\frac{\wt{\Ex{r^2}}}{\Ex{r^2}}\simeq s^2.
$$
Combining these results yields the following forms for the reactant density in two dimensions
\b
n(t)=\frac{\log\Ex{r^2}}{6\pi \Ex{r^2}}+O\left(\Ex{r^2}^{-1}\right)&\hspace{0.2cm}&n(t)=\frac{\log(t)}{24\pi D_Rt}+O\left(t^{-1}\right). \label{2D} 
\e
where in the second expression the density has been rewritten in terms of time by using the result (\ref{DL}). At this point comparison can be made with the MARW. The result (\ref{PDC}) and (\ref{2D}) are of the same form, but differ from each other by the amplitude: the disorder renormalisation of the reaction term has decreased it from $1/8\pi$ to $1/24\pi$. This implies that reactions occur at an increased rate in the presence of Sinai disorder.

It is interesting at this point to consider the behaviour of different lattice models with fixed diffusion constant $D$. By writing a CS equation for the diffusion length the following relation between the effective diffusion constant measured at late time $D_R$, and the lattice-model parameter $D$ can be obtained
\b
D_R&=&D\left(1-\frac{\Delta}{2\pi D^2}+O\left(\Delta^2\right)\right). \label{DREL}
\e
The above relation is valid for weak disorder (small $\Delta$) and implies that the effective diffusion constant is reduced from the lattice-model value $D$ by the disorder strength. Hence, if written in terms of the long time and length scale behaviour of a lattice model with parameters $D$ and $\Delta$, the reactant density becomes
\b
&&n(t)=\frac{\log(t)}{24\pi Dt}\left(1+\frac{\Delta}{2\pi D^2}\right)+O\left(t^{-1}\right)+O\left(\Delta^2\right). \label{2Dbare} 
\e
In terms of the lattice parameter $D$ it is seen that the reaction rate is still faster than the MARW. However, as the disorder strength is increased the reaction rate starts to decrease. Unfortunately, the result is valid only for weak disorder and it is not possible to determine from (\ref{2Dbare}) if a point is reached where there is a cross-over and the reaction rate becomes less than the MARW.

\section{Discussion}
We have examined the late-time density of reactants in the single-species process $A+A\rightarrow O$ in the presence of an uncorrelated, quenched random velocity field: so called Sinai disorder. Contrary to many existing works on reactions in disorder, the statistics chosen for this velocity field were such that there were {\it no long-range correlations}. Despite the lack of correlations, it was shown that in two dimensions and below disorder changes the density decay on all time scales. The model, introduced in section (2), was analysed in the language of field theory and the renormalisation group was used to obtain the late-time reactant density. It was shown that the appropriate action that generates the disorder-averaged correlation functions is a combination of terms seen in the field-theoretic analyses of diffusion in Sinai disorder \cite{F1} and the $A+A\rightarrow O$ reaction in the absence of disorder \cite{BPL}. The new interaction diagrams that appear in this hybridised theory were identified up to the one-loop level, and the perturbative density was calculated. It was shown that at this level a new term appears corresponding to an annihilation of two particles that both interacted with the same region of the random field at earlier times. In section (3) the late-time forms for the reactant density were derived from the perturbative results, by the use of the appropriate Callan-Symanzik equation (\ref{CS}). 

Below two dimensions the effects of trapping in Sinai disorder are severe and it was found that the reaction process becomes {\it sub-diffusion limited}, equation (\ref{NT}). It was shown that because of the changed dynamic exponent, the relation between the reactant density and time must be non-universal, i.e. dependent on the reaction rate or disorder strength. By writing the density as a function of time, the pure diffusion length $Dt$ is the implicit length scale. However, in the presence of disorder this is an inappropriate scale. Rather, the density should be written as a function of the disorder-averaged diffusion-length squared $\Ex{r^2}$. By exchanging the dependence in $t$ for $\Ex{r^2}$ at the level of the CS equation a universal relation between these two non-universal quantities was obtained that is independent of the reaction rate, disorder strength and initial density.

At the upper-critical dimension $d_c=2$ the asymptotically-exact form of the density was obtained as a function of time, equation (\ref{2D}). The random velocity field has the effect of reducing the amplitude of the density decay which  implies that for weak disorder the rate of reaction is {\it faster} than for purely diffusing reactants: an effect coming from the disorder renormalisation of the reaction term given in figure (1c). Physically, it represents the process whereby two reactants are pushed into the same region of space by the disorder and therefore brought closer together than if they were simply diffusing without a bias. Taking the view-point of the lattice model, it is appropriate to express the density decay in terms of the model parameter $D$ rather than the measured, late-time effective diffusion constant $D_R$. In this case, it is seen that as the disorder strength is increased the reaction rate begins to {\it decrease}, equation (\ref{2Dbare}). This occurs because the rate at which the particles explore space is reduced due to the diffusion-constant renormalisation, an effect coming from the dressing of the propagator by the disorder vertex. The result obtained here is an expansion in the disorder strength $\Delta$ and is therefore only valid only for weak disorder. It would be interesting to obtain results for the strong disorder case, perhaps from a numerical approach, to see if increasing the disorder further produces a density decay {\it slower} than the MARW.

Finally, we briefly compare the effects of the disorder examined in this paper and the case of long-ranged {\it potential} disorder \cite{AAPD} \cite{CD}. The effect of potential disorder in two dimensions is more drastic because the exponent of the decay is changed, whereas for Sinai disorder the amplitude is altered. The relative severity can be understood by the nature of the disordered landscape. In a study of diffusion in various forms of disordered landscapes \cite{KLY}, it was noted that potential disorder produces a landscape with deep trapping wells where, to escape from a trap, any path a particle might take involves movement in an unfavourable direction. However, this is not the case for Sinai disorder where the landscape (in two dimensions) does not have the morphology of potential wells and any pseudo-traps that might exist will tend to have velocity drifts nearby that allow for escape.

\nsubsecnn{Acknowledgements}
We would like to thank Dr M. W. Deem,  Dr J.-M. Park and Dr G. M. Sch\"utz for useful discussions. Prof. D. Mukamel and Y. Kafri are thanked for useful comments on the manuscript . This research was partly supported by the
Engineering and Physical Sciences Research Council under Grant
GR/J78327.

\end{document}